# Remembering Wassily Hoeffding

**Nicholas I. Fisher and Willem R. van Zwet**

*Abstract.* Wasssily Hoeffding's terminal illness and untimely death in 1991 put an end to efforts that were made to interview him for *Statistical Science*. An account of his scientific work is given in Fisher and Sen [*The Collected Works of Wassily Hoeffding* (1994) Springer], but the present authors felt that the statistical community should also be told about the life of this remarkable man. He contributed much to statistical science, but will also live on in the memory of those who knew him as a kind and modest teacher and friend, whose courage and learning were matched by a wonderful sense of humor.

*Key words and phrases:* Asymptotic optimality, copula, Hoeffding's decomposition, Hoeffding Race, likelihood ratio test for multinomial, local alternatives, locally most powerful rank tests, $m$-dependence, permutation tests, probability inequalities, random permutations, rank correlation, statistical correlation, $U$-statistics.

## INTRODUCTION

Wassily Hoeffding was one of the founding fathers of nonparametric statistics, the science of analyzing data without making unnecessarily restrictive assumptions about their origin. His great strength was his deep understanding of statistics that told him which problems to attack, at what level of generality and with what mathematical means. He was not interested in generality for generality's sake and he always kept his mathematics as simple as possible. When he had dealt with a problem, he left it to others to examine the consequences of his results. It is striking how often he picked the "right" problems, the ones that much later turned out to have consequences that went far beyond his deceptively simple results, and opened up entirely new fields of research. He was the statistician's statistician in the sense that he supplied what his colleagues needed today or would need in future.

Wassily was the most unassuming person imaginable with a mild but permanently present sense of humor. Throughout his life he suffered from diabetes, bad eyesight and hearing, but his indomitable spirit made these handicaps seem like minor inconveniences. As our colleague Ildar Ibragimov wrote[1] to us from Saint Petersburg, Wassily Hoeffding was a very intelligent and broadly educated person with lively interests and a truly noble soul that we rarely come across even in great people.

## THE EARLY YEARS, 1914–1947

Wassily was born in Mustamäki, Finland (Gorkovskoye, USSR since 1940), although his place of birth is registered as Saint Petersburg on his birth certificate. His parents had traveled from their home in Tsarskoye Selo (now Pushkin) to spend the summer in Mustamäki. In his autobiographical article (Hoeffding, 1982, page 100), Wassily provided some family history:

*Nicholas I. Fisher is Visiting Professor of Statistics, School of Mathematics & Statistics F07, University of Sydney, NSW 2006, Australia e-mail: nickf@maths.usyd.edu.au. Willem R. van Zwet is Emeritus Professor of Statistics, Mathematical Institute, University of Leiden, P.O. Box 9512, 2300 RA Leiden, The Netherlands e-mail: vanzwet@math. leidenuniv.nl.*

---

[1]Email message from Ildar Ibragimov dated 25 November, 1997.

My father, whose parents were Danish, was an economist and a disciple of Peter Struve, the Russian social scientist and public figure. An uncle of my father's was Harald Hoeffding,[2] the philosopher. My mother, née Wedensky, had studied medicine. Both grandfathers had been engineers.

His elder brother Waldemar became a doctor; he immigrated to the United States before World War II. His younger brother Oleg[3] became a military historian in the U.S., and Oleg's life interleaved in interesting ways with Wassily's, as we shall see.

In 1918, the family left Tsarskoye Selo for Ukraine and, after traveling through scenes of civil war, finally left Russia for Denmark in 1920, where he entered school.

Hoeffding wrote:

> In 1924 the family settled in Berlin. In high school, an Oberrealschule which put emphasis on natural sciences and modern languages, I liked mathematics and biology and disliked physics. When I finished high school in 1933, I had no definite career in mind. I thought I would become an economist like my father and entered the Handelschochschule... in Berlin.

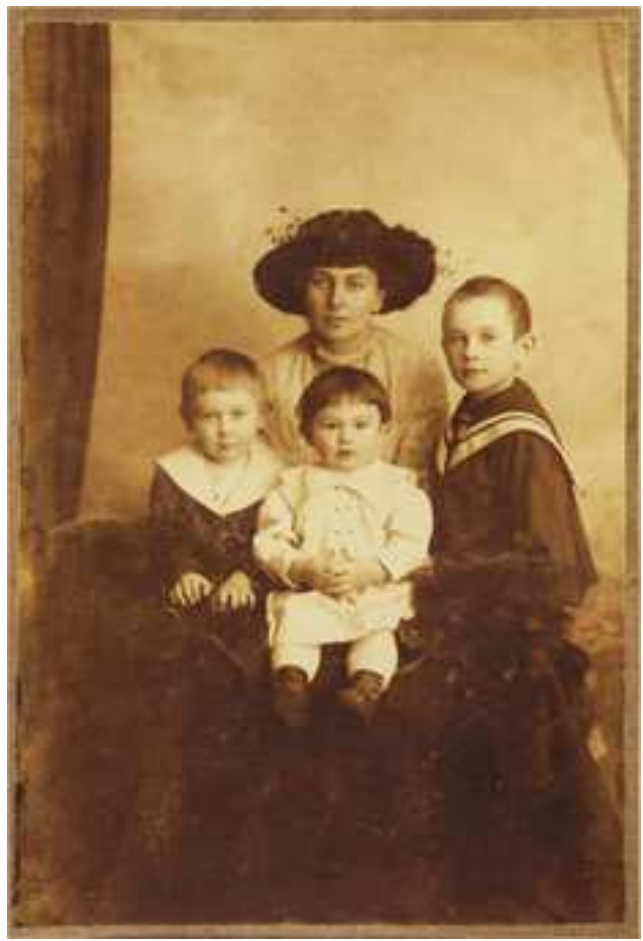

FIG. 1. *WH with his mother and two brothers, Waldemar (the elder) and Oleg (the younger).*

> But I soon found that economics was too vague a science for me. Chance phenomena and their laws captured my interest. I performed series of random tossings and recorded their outcomes before I knew much about probability theory. One of the few books on chance phenomena that I found in the library of the Hochschule was *Die Analyse des Zufalls* by H.E. Timerding, and it fascinated me. In 1934 I entered Berlin University to study mathematics... The meager fare in mathematical statistics that I was fed in my lectures in Berlin, I tried to supplement by reading journals. But somehow I did not fully absorb the spirit of research at the frontier of the subject in my student days. My Ph.D. dissertation... was in descriptive statistics and did not deal with sampling. ...My Doktorvater or Ph.D. supervisor was Klose.

---

[2]The Columbia Encyclopedia (Sixth Edition, 2001) records the following information: Harald Høffding 1843–1931, Danish philosopher. He was professor at Copenhagen (1883–1915). His histories of philosophy have been enjoyed by a large audience, especially his History of Modern Philosophy (1894–95; tr., 2 vol., 1900, reprinted by Dover 1955).

[3]In an email message dated 23 January 2002, Oleg's daughter, Virginia Hoeffding, wrote: "Oleg spent the war in England, and wound up attached to U.S. Army Intelligence during the liberation of Europe; among other things, he was the first Allied intelligence officer to interview Albert Speer (he tends to appear in biographies of Speer as Captain "Otto" Hoeffding—much to his annoyance). He and my mother and brother came to this country in 1946... the American intelligence community felt he was a valuable resource, and essentially sponsored his hiring by Columbia University as an instructor in the Economics Department. ...He left Columbia for the Rand Corporation when the latter was founded in 1953, and remained there until his retirement in the mid '70s. His field was Sino-Soviet economic relations, so that a great deal of his work was and remains classified. He was one of several people associated with Daniel Ellsberg during the Pentagon Papers episode, and was one of the co-signers of Ellsberg's famous letter to *The New York Times*. His opposition to the war was, I think, largely pragmatic rather than ideological; one of his papers which is in the public domain was an analysis of the ineffectiveness of strategic bombing."

I chose the topic of the thesis and worked on it largely by myself, with some suggestions and encouragement from him. He was a Baltic German and had his own ideas about Russians. He warned me to refrain from making exaggerated claims in my thesis that I could not substantiate as, he thought, Russians were prone to do. (*Ibid.*, page 100.)

Wassily's first published papers (e.g., Hoeffding, 1940) appeared in a Series of the University of Berlin, and reported (in German) the results of his Ph.D. investigation into the correlation phenomenon. Characteristically, he at once saw the "right" formulation of this problem. He studied correlation properties of bivariate distributions that are invariant under arbitrary monotone transformations of the marginals. The work is related to later work on dependence by Fréchet and Lehmann. But the concept of invariance under monotone transformations is relevant in a much broader setting and lies at the root of nonparametric or rank methods, the major development in statistics in the years following World War II. In fact, in Hoeffding's correlation papers one encounters Spearman's rank correlation coefficient, which also predates the formal development of rank statistics. It is not surprising that his first published work in an international journal deals with rank correlation (Hoeffding, 1947).

The war, and the relative obscurity of University Series to statistical researchers elsewhere, meant that his results were known only to the few people outside Germany to whom he had sent copies of his thesis. It is only recently, with the publication of English translations of these papers in his Collected Works (Fisher and Sen, 1994), that these early results, rediscovered many years later by others, are generally acknowledged to be Hoeffding's.

However, he did not proceed immediately to a research career.

On completing my studies in 1940, I accepted two part-time jobs: as an editorial assistant with the *Jahrbuch über die Fortschritte der Mathematik* and as a research assistant with the interuniversity institute for actuarial science... I held both jobs until almost the end of the war. I never applied for a teaching position in Germany: I had been stateless since leaving Russia and did not wish to acquire German citizenship, which was necessary to hold a university teaching job. (*Ibid.*, page 101–102.)

Meanwhile, brother Oleg had spent the war in England, working in the Economic Division of the U.S. Embassy. Towards the end of the war, Wassily and his mother experienced a very heavy bombing raid. He told Ildar Ibragimov of discovering, subsequently, that Oleg had been involved in planning the raid.[4]

In February 1945 I left Berlin with my mother for a small town in the province of Hanover to stay with a Swiss friend of my father's... My father stayed behind and was captured by what was later to become the KGB. He had been employed for many years at the office of the American Commercial Attaché and then had been the economic correspondent of American and Swiss periodicals. This made him a "spy" in the eyes of the KGB. (Ibid., page 103.)

During this time, Wassily asked Oleg to send him a copy of M. G. Kendall's recently released Volume 1 of *The Advanced Theory of Statistics*. "It read to me like a revelation," he wrote. And it was also in Hanover that he wrote "my first statistical paper in the modern sense of the word," establishing the asymptotic normality of Kendall's rank correlation coefficient $\tau$ in the general case of independent identically distributed random vectors. However, the significance of the paper resided not so much in its content, but in the path down which it led him: to his ground-breaking theory of $U$-statistics. But first, he had to leave Germany and find a job.

We left Germany for Switzerland and arrived in New York City in September 1946... As I was unemployed, I attended lectures at Columbia University by Abraham Wald, Jack Wolfowitz and Jerzy Neyman, who was then visiting Columbia. I was in the thick of contemporary statistics... In 1940 I had sent a few copies of my Ph.D. thesis to statisticians in other countries, including the United States... Thus my name was not entirely unknown in the USA. My brother Oleg, whose arrival in New York preceded mine by three months and who now was an economics instructor at Columbia, helped me to get invitations from the Cowles Commission for Economic Research... and from Harold Hotelling, who had just established the Department of Mathematical Statistics at the University of North Carolina at Chapel Hill.
I first went to Chicago to give a talk on what I later called U-statistics... Soon after, a letter from Hotelling offered me a position as research associate. He did not ask for a preliminary visit; later he said he had been impressed that a Ph.D. thesis in mathematical statistics had come out of Germany. In May 1947 I arrived in Chapel Hill. (Ibid., page 103–104.)

---

[4]Hanover became part of the British zone of occupation, and Wassily and his mother stayed there for over a year, trying in vain to secure the release of his father. Then, they lost all trace of him. Some time after they had left Germany, Hoeffding *père* escaped from prison in Potsdam.

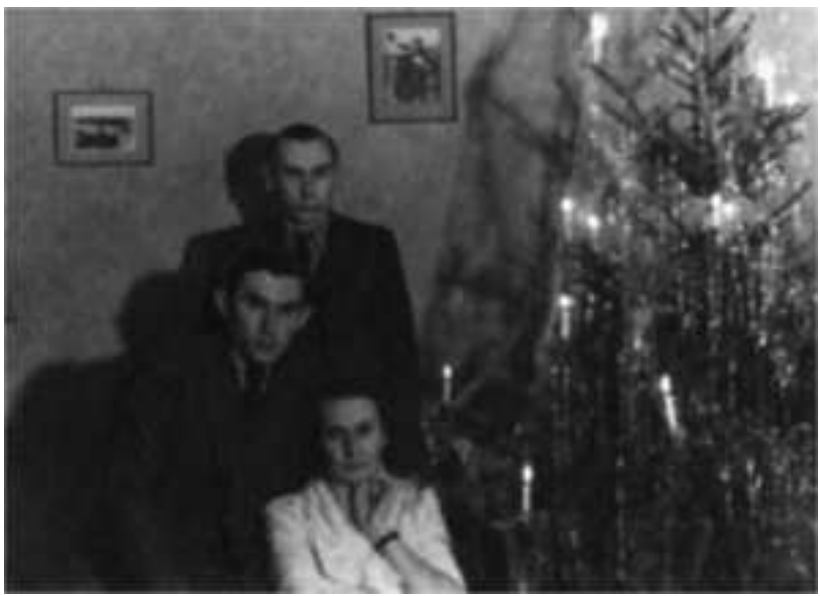

Fig. 2. *WH with his mother and his brother Oleg, Christmas, possibly taken before the war.*

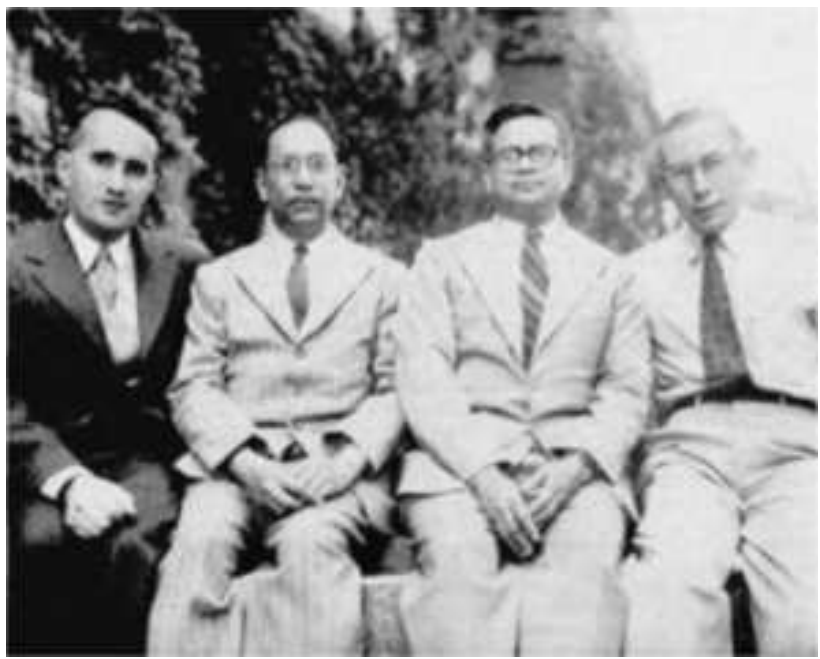

Fig. 3. *WH with colleagues a few years after he arrived in Chapel Hill. The photograph was supplied by Walter Smith from the "Record of Research" of the Institute of Statistics, Volume II, July 1951 to June 30 1953, probably taken in the spring of 1953 when NLJ was visiting Chapel Hill. The photo is incomplete in that it does not show Harold Hotelling or George Nicholson (who was chairman at this time).*

Wassily had arrived at his academic home. Apart from conference travel, and the occasional short visit (six months or less) to other scientific institutions, he remained in Chapel Hill for the rest of his life.

I was to remain in Chapel Hill until my retirement and beyond. Congenial colleagues, a relaxed, informal academic life style, the attractive nature of the town, the relative closeness of the sea and the mountains, combined with an inborn inertia, made me resist the temptations of moving to other campuses. Being somewhat reserved by nature, I cherish all the more the friendships and contacts I have had with my colleagues and students in the department and their families. (*Ibid.*, page 105.)

## ACADEMIC CAREER AT UNC-CHAPEL HILL, 1947–1979

The Department of Statistics at Chapel Hill indeed provided a congenial professional environment. Hotelling had attracted several leading researchers, including R. C. Bose, S. N. Roy, P. L. Hsu, E. J. G. Pitman and Herbert Robbins. Wassily immediately embarked on an active research career. A year after his arrival, he published a paper on $U$-statistics (Hoeffding, 1948) that has become a landmark in the development of asymptotic statistics. A $U$-statistic is of the form

$$U_n = \Sigma \cdots \Sigma h(X_{i(1)}, \ldots, X_{i(k)}),$$

where $X_1, \ldots, X_n$ are independent and identically distributed (i.i.d.) random variables and the summation extends over all $k$-tuples of distinct elements of $\{1, \ldots, n\}$. $U$-statistics were introduced by Halmos as unbiased estimators of their expectation, but are found to play a role in almost any statistical setting. From a probabilistic point of view, $U$-statistics of degree $k = 2, 3, \ldots$ are successive generalizations of sums of i.i.d. random variables (the case $k = 1$), the study of which has formed the central part of probability theory for centuries. Hoeffding performed the essential moment calculations for $U$-statistics, established their consistency as well as their asymptotic normality as $n \to \infty$, and dealt with the case where the $X_i$ are not identically distributed as an encore. Wassily liked to think of this paper as his "real" Ph.D. dissertation. It is fair to say that this paper started and the same time finished the probabilistic study of $U$-statistics until Hoeffding himself took up the subject once more some 13 years later.

In an important unpublished paper (Hoeffding, 1961), which is discussed but not reprinted in his *Collected Works*, he introduced a decomposition of an arbitrary function of $X_1, \ldots, X_n$ in terms of $U$-statistics. This is now known as Hoeffding's decomposition and has proved to be a powerful tool in asymptotic statistical theory. The leading term of the expansion was extensively used by Hajek for proving asymptotic normality and became known as Hajek's projection.

At the time Wassily was sharing an office with Herbert Robbins and they soon became close friends. They collaborated on a paper (Hoeffding and Robbins, 1948) on a problem of common interest, the central limit theorem for $m$-dependent variables. For Wassily there was an obvious link with his work on $U$-statistics and the paper may be considered as a starting point for the vast literature on the central limit theorem under mixing conditions. Three years later Hoeffding returned to the central limit theorem under non-standard conditions when he proved a combinatorial central limit theorem (Hoeffding, 1951a): if $(R_1, \ldots, R_n)$ is a random permutation of the numbers $1, 2, \ldots, n$ that assumes every permutation with the same probability $1/n!$, then

(1) $$T_n = \Sigma_j b(j, R_j)$$

is asymptotically normal as $n \to \infty$ under simple conditions which were later proved to be minimal. This result has turned out to be the workhorse of the asymptotic theory of rank tests which often have statistics that are either of the form (1), or that can be approximated by such statistics.

Having contributed one of the major tools for studying of rank tests, it was time for Hoeffding to take part in this study. His first paper in this area (Hoeffding, 1951b) was a major one. The appealing property of rank tests is that the probability of an error of the first kind—that is, rejecting the null-hypothesis when it is true—is known and constant whenever the hypothesis is satisfied. However, the power of the test—that is, the probability of rejecting the hypothesis if the alternative is true—is generally unmanageable even for parametric alternatives. Except in very simple cases it is not possible to construct a rank test that is most powerful for a specific parametric alternative. To get around this problem Hoeffding introduced the concept of a locally most powerful rank test, that is a rank test which is most powerful in a typically parametric arbitrarily small neighborhood of the null hypothesis. Hoeffding showed that in many cases such tests could be easily derived and often performed quite well in general. But the meaning of this result was much deeper than that. It later became clear that asymptotically as $n \to \infty$, these local alternatives are the only ones that are relevant for the asymptotic power of a test, as a consequence of which a direct link was established between locally most powerful and asymptotically optimal rank tests. The way to an asymptotic optimality theory was clearly mapped out and Wassily would expound his ideas on asymptotic efficiency of tests three years later in a joint paper with Joan Rosenblatt (Hoeffding and Rosenblatt, 1955).

Permutation tests share the property of rank tests that the probability of an error of the first kind is constant whenever the hypothesis is satisfied. They are typically modelled after parametric tests performed conditionally on some aspect of the data that is irrelevant for the truth or falsehood of the hypothesis to be tested. With the techniques available at the time, analysing the performance of nonparametric tests in general was not yet possible, but Hoeffding achieved the *tour de force* of showing that permutation tests were asymptotically as powerful in the parametric setting as the parametric tests on which they were modelled (Hoeffding, 1952).

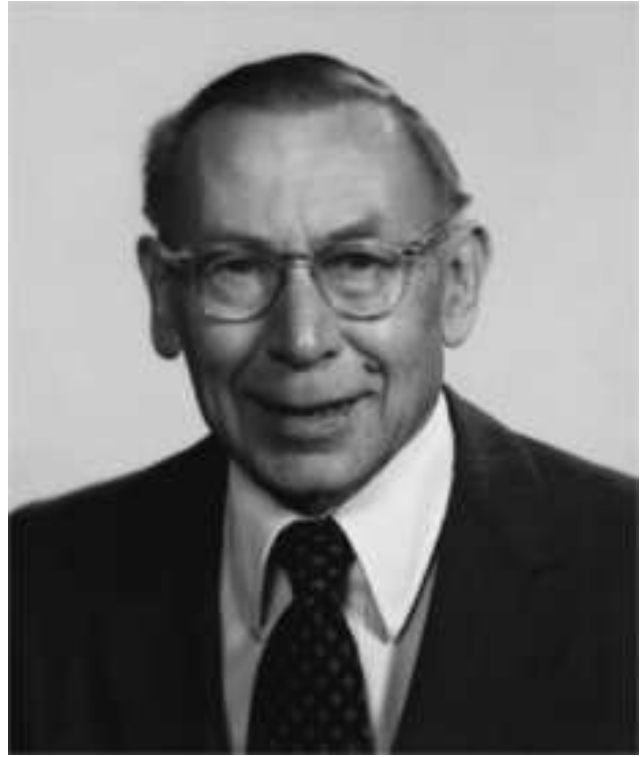

FIG. 4. *WH portrait taken in late '60s or early '70s.*

There followed a remarkable series of papers concerning bounds on expected values of functions of sums of independent random variables, of which Hoeffding (1956) generated most attention. For independent random variables $X_1, \ldots, X_n$, let $X_j$ assume the values 0 and 1 with probabilities $p_j$ and $(1-p_j)$ respectively, and let $\Sigma p_j = np$ for given $p \in (0,1)$. It is shown that according to several criteria, $\Sigma X_j$ is most spread out if all $p_j = p$. In other words, the binomial distribution is the least concentrated among all Poisson-binomial distributions with the same expectation. This result has many intriguing ramifications.

Hoeffding (1963) established probability inequalities for sums of bounded random variables. This paper has been truly influential in many areas of statistics including empirical process theory. Perhaps more surprisingly, it is one of the most cited papers in the computer science literature; for example, it is used in algorithms simulating pinball machines, and is associated with the term "a Hoeffding Race." Hoeffding (1965, 1967) are devoted to probabilities of large deviations and a related optimality property of likelihood ratio tests for multinomial distributions.

An extensive discussion of the full breadth and impact of Wassily's scientific contributions is given in four papers by some of his colleagues, included in his *Collected Works.*

## WASSILY AS TEACHER

As with all other aspects of his professional life, Wassily approached his interactions with students with very careful thought. In his later years in the department, his teaching was confined to advanced graduate courses,[5] of which he taught five: Estimation and Hypothesis Testing, Nonparametric Statistics, Sequential Analysis, Decision Theory, and Large-Sample Theory. For this level of student, the courses were a delight: as elegantly and meticulously prepared as his research papers, always totally up to date and above all designed to promote understanding and learning. Further, his examinations were a revelation. He took the view that students taking his courses wanted to learn, and that it wasn't his job to seek to catch them out by probing for areas of confusion or lack of knowledge. Rather, his examinations, which took the form of one-week take-home projects, simply extended the learning process, by leading the student through an area of the subject not covered in the course. For example, on one occasion he chose not to cover $U$-statistics in presenting his Nonparametric Statistics course, but the take-home exam resulted in the students becoming familiar with the area. Of course, when he taught the course on another occasion, a different topic would be omitted.

One delightful classroom story was handed down from one generation of students to the next. There tended to be little dialog in Wassily's lectures; it was mainly a question of copying the material that he wrote on the board in chalk. However, legend had it that on one occasion a student interrupted him to ask, "Excuse me, Professor Hoeffding, but can you give us an example of this theorem, please?" Wassily thought briefly and then wrote on the board,

$$\text{Example: set } \theta = \theta_0.$$

---

[5]With one outstanding exception: on one occasion, he was assigned (by some unknown and unfathomable process) to teach an introductory service course in statistics to a class of undergraduates comprising basketball players, footballers and other assorted attendees, none of whom was attending UNC-CH with the primary intent of learning statistics. It is not clear who received the greater shock from the semester's experience: the students or Wassily.

Wassily supervised the Ph.D. theses of 17 students. As an advisor, he was supportive, but nonintrusive. Indeed, his natural modesty made the early phase of Ph.D. research somewhat daunting. When seeking advice about the current difficulty with which one was confronted, it was essential to end a description of the problem with a direct question. Otherwise (and typically) Wassily did not presume that his counsel was being sought and so would offer no comment.

## LIFE OUTSIDE STATISTICS

Although not disposed to lengthy sojourns away from Chapel Hill, Wassily very much enjoyed the shorter trips that he made to other parts of America and the world, usually in association with his work. An early Berkeley Symposium provided him with the opportunity to meet statisticians and probabilists from the West Coast and from all over the world. He recorded his delight at the opportunity to go for a three-day hike in Yosemite Park with some colleagues. Walking with colleagues in the hills around Ithaca was a highlight in a later visit to Cornell. Attendance at international meetings took him to Sydney, Warsaw, Stockholm and other foreign parts; a six-month stint in India, funded by UNESCO, included one month of touring around the country, giving lectures in Benares, Lucknow, Delhi, Agra, Bombay, Bangalore, Mysore City, Trivandrum and Madras. After India, the U.S. National Academy of Sciences sponsored a one-month visit to Russia, where he visited Moscow, Leningrad, Kiev, Tashkent and Novosibirsk. Subsequently, while attending a conference in Warsaw he was able to visit Vilnius.

> The visit to Akademgorodok, the seat of the Siberian section of the Academy of Sciences of the USSR, near Novosibirsk, was of special interest. Few Westerners had been there before me. I was warmly received by A. A. Borovkov and his co-workers, mostly younger people. I had arrived directly from Tashkent, where the flowers were in bloom in the public squares. Here, when we were walking on the ice-bound Ob River in a chilly wind, I put on the only head covering I had with me, an embroidered Uzbek scull cap which had been presented to me in Tashkent. (Ibid., page 108.)

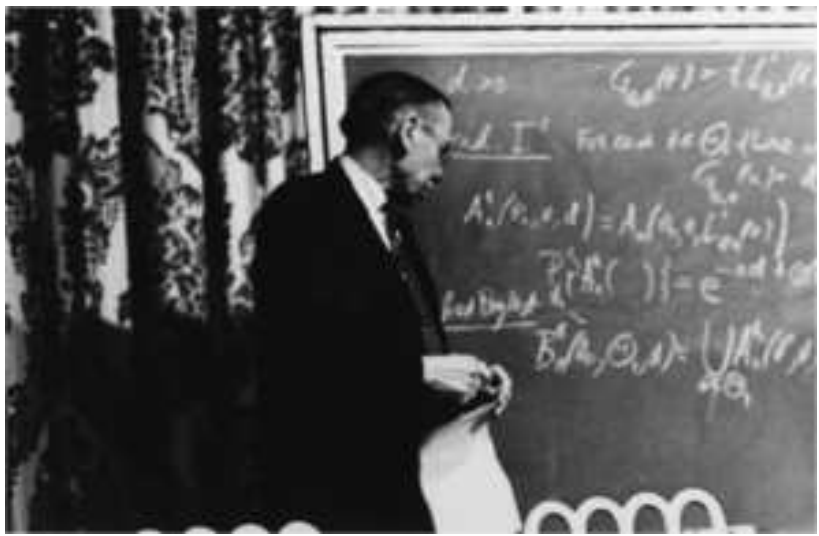

Fig. 5. *WH presenting one of the Wald Lectures in Washington DC, September 1967. (Photograph by Charles Roberts.)*

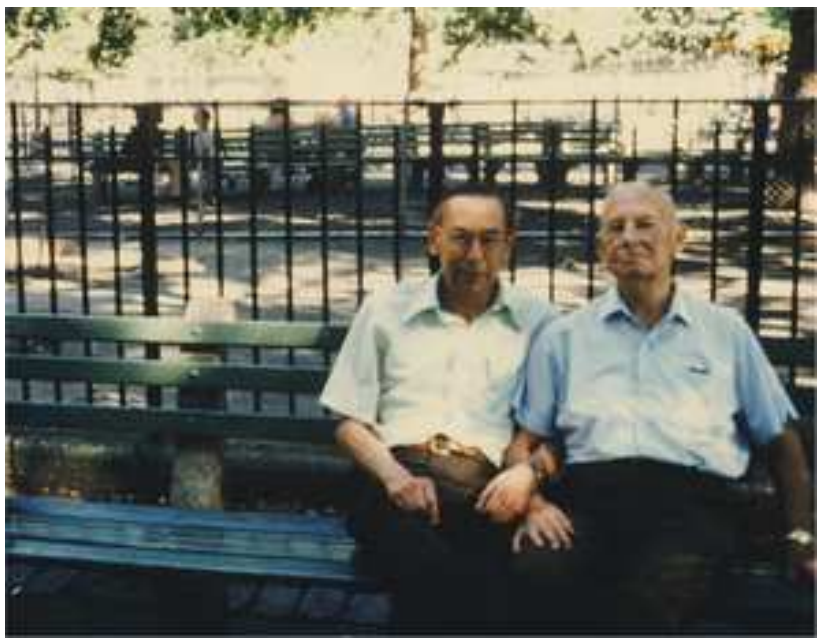

Fig. 6. *WH with his older brother, Waldemar, at Battery Park, Chapel Hill, 10 August 1974.*

Apart from Wassily's enjoyment of hiking, his long-time friend and colleague, Ross Leadbetter, records[6] that

> ...operating his sailboat brought him much pleasure, even though as he liked to recall, he was once arrested by the coastguard out in Kerr Lake for the heinous crime of having non-standard life jackets. A rather amusing train of correspondence ensured, which it appears the authorities tired of first, dropping the matter.

In fact, this was no ordinary boat...

> At a department lunch party at the Carolina Inn, Wassily was seated next to a young secretary. Being less comfortable than

---

[6]Memorial Service for Wassily Hoeffding, March 7, 1991, Person Recital Hall, University of North Carolina at Chapel Hill.

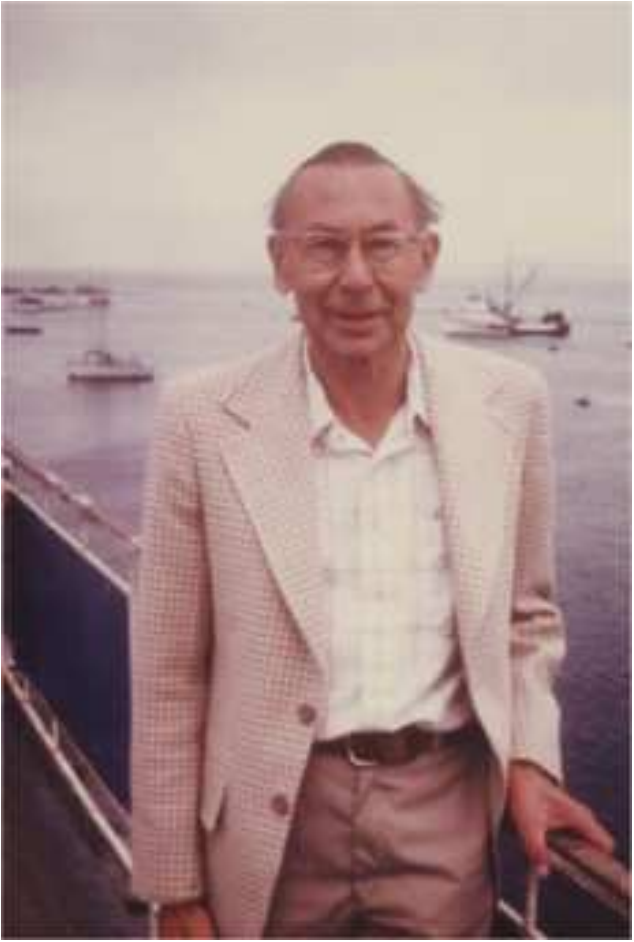

Fig. 7. *WH at Monterey, California, July 1977.*

he was with no conversation she tried to draw him out, with the following result:

    Secretary: "Prof. Hoeffding, I understand you have a boat."
    Wassily: "Yes." [*Silence*]
A little later she made another attempt:
    Secretary: "Prof. Hoeffding, where do you keep your boat?"
    Wassily: "In my closet."
End of conversation.

It was actually quite a fancy boat that could be disassembled and stored in his closet.

He was always well informed on world affairs, and was deeply interested in cultural matters. Ildar Ibragimov wrote[7] to us of Wassily's interest in Russian literature, particularly Russian poetry. Wassily used to send Ibragimov Western editions—in Russian—of the Russian poets Akmatova and Gunilev and Ibragimov subsequently sent him very good Russian editions of these poets. On one occasion when Wassily and his mother visited Tashkent, they were invited to a dinner. Someone proposed that they all toast their motherlands: USSR and USA. Wassily and his mother stood up and proclaimed the famous lines from Pushkin:

> "Nam celyi mir chuzhbina
> Otechestvo nam Carskoe Selo"
> ("For the whole world is a strange country,
> Our motherland is Tsarskoe Selo")

Tsarskoe Selo was, of course, Wassily's home town.

## PROFESSIONAL HONORS

Wassily was the recipient of the many of the highest honors available to statisticians, in the U.S. and overseas. These included:

1967 Wald Lecturer
1969 President, Institute of Mathematical Statistics
1973 Appointed Kenan Professor, University of North Carolina
1976 Elected to National Academy of Sciences
1985 Elected to American Academy of Arts and Sciences
Elected Fellow of the American Statistical Association
Elected Fellow of the Institute of Mathematical Statistics
Elected Member of the International Statistical Institute
Elected Honorary Fellow of the Royal Statistical Society

Shortly after his retirement at the age of 65, the University of North Carolina's College of Arts and Sciences established the Wassily Hoeffding Professorship in his honor.

## RETIREMENT, 1979–1991

In reflecting on his career, Wassily concluded:

> Ever since I switched from economics to probability and statistics in my early student days, this area has continued to absorb my interests. The very idea that the seeming chaos of chance obeys mathematical laws is immensely attractive. It gives

---

[7]Email message from Ildar Ibragimov dated 25 November, 1997.

me great satisfaction to have made a few contributions to the understanding of this field.

The successes I had did not come easy to me. They were the fruits of long hours of work which often led to dead ends. I am well aware that with advancing years my capacity to work has diminished. The lure of the subject persists. Whether I will contribute more to it, only time will tell. (Ibid., page 108.)

Just before his 65th birthday, an international symposium was held in his honor (Chakravarti, 1980). Sadly, he was suddenly taken ill during the symposium dinner and rushed to hospital, where

> ...my right leg had to be amputated. (The reason was an infection related to my diabetes.) Since then, I have been getting used to a new kind of life. (Ibid., page 108.)

He remained active and alert. His new life still involved research, notably a number of typically elegant contributions to the Encyclopedia of Statistical Sciences. In describing Wassily's retirement activities, Ross Leadbetter commented at the memorial service:

> "Those of us privileged to enjoy his friendship in later years must be struck by his extraordinary strength of character, his extreme generosity—almost to a fault—towards causes, and people expressing needs, and his amazing ability to contend with the seemingly endless medical complications, not too infrequently life threatening, and yet to get very significant enjoyment from life. His later life was burdened severely by the necessities of medical attention but he would make the fullest use of intervening time with avid reading, with writing made laborious by poor circulation in his fingers, watching TV and listening to his shortwave radio. He would scan the Annals of Statistics for any articles involving $U$-statistics. He was fond of Russian literature and poetry and was busy reading a new and extensive biography of Tolstoy during his recent stay in hospital. A New York Times reader without peer, he took endless delight in finding the unusual new items which appealed to a very real sense of humor that could surprise those only casually acquainted with him. He clearly enjoyed his later years in spite of the struggles and limitations. He enjoyed his team of nurses, his watchful doctor..., his neighbors..., his local friends and international department visitors who would want to go and see him. He kept his dignity to the end, and never gave any hint of self pity. As a basically theoretical person he gave many practical lessons to those about him, not the least of which was that real friendship is something beyond the clatter of small talk, and is better demonstrated by action rather than endless words."

Tributes from friends, colleagues and even from those who had never met him, provide a clear and consistent picture of this remarkable human being. In a Memorial Session held at Joint Annual Statistical Meetings in Boston, August 1991, Richard Vitale, who fell into this last category, said:

> I never had the opportunity to meet Professor Hoeffding personally but am glad to have encountered him through his work. Several times in conducting a literature search, I found that he was the one who pioneered a certain problem area. He always did it with a penetrating understanding of the fundamental questions and elegant, lucid analysis. Or, to be more direct, he always seemed to understand and do a problem right, the way it *ought* to be done. He was for me, in this way, a gifted and generous teacher.

In the same session a tribute was read from his collaborator from his early days, Herbert Robbins:

> Although he was gentle and courteous in manner and fragile in health, he was lionhearted in spirit and completely original in his scientific work. His character was truly noble; I never heard from him a complaint about his chronic illness or the difficulties of living in an alien environment, or a disparaging remark about any fellow human being. The Statistics Departments of Columbia and Rutgers join in expressing their sorrow at the death of Wassily Hoeffding, one of the great creative forces of statistical thought of our time. His work will long continue to provide an unequalled example of mathematical elegance and manifold application. His presence here from 1946 until his death has greatly enriched the intellectual and cultural life of this country.

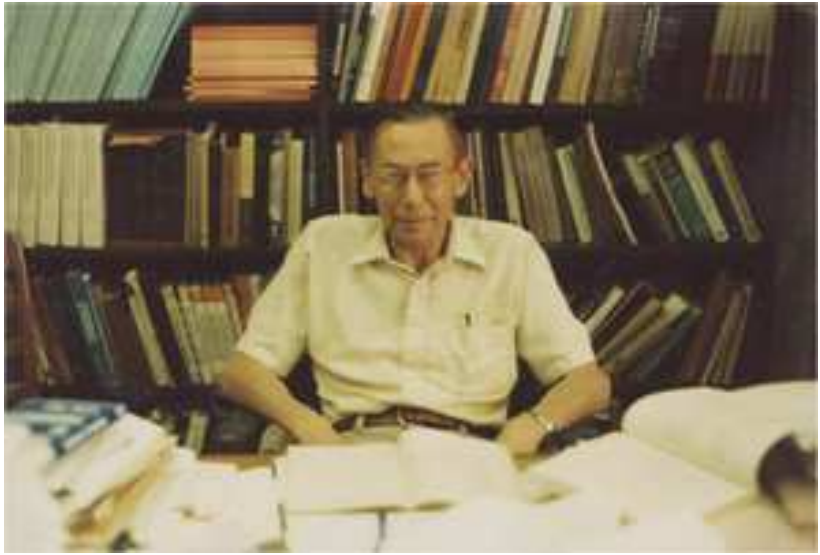

Fig. 8. *WH as many of his colleagues, friends and students at Chapel Hill remember him, in his office in Phillips Hall, University of North Carolina at Chapel Hill. (Photograph probably taken late '60s or early '70s.)*

Bill Cleveland recalled social interactions from around 1970, at UNC:

> The commons room for the Statistics Department in Phillips Hall at the University of North Carolina, Chapel Hill, was a social place where faculty met, and talked, occasionally about statistics but typically about life in general. It was in the commons room that I got to know Wassily: smart, formal but not stuffy, and a good sense of humor. Very likeable. One day, he and I and Wally Smith were chatting, and talk turned to World War II and Wassily's job in Germany as an airplane spotter. Suddenly, much to my amazement, Wally, who had been Wassily's colleague for years, asked: "Wassily, who did you want to win the war?" A look of incredulity spread across Wassily's face, and in a tone of voice that indicated there was one, obvious answer he said: "Well, the Allies, of course." Wassily had a charming candor. And he was not timid behind the wheel of a car. Both came together one day. A police officer pulled him over and told him he was driving 70 mph in a 50 mph zone. Wassily, distressed, said to the officer: "That is terrible. I know the speed limit is 50 on this road, so normally I try to drive at 60." The police office was very amused and gave Wassily a ticket for 60 mph.

Long-time colleague Gordon Simons related his experience in collaborating with Wassily:

> Wassily had a playful spirit, which eventually led to a cute [joint] paper (Hoeffding and Simons, 1970). ... At the time, Wassily was president of the IMS and had the task of making various appointments. As I recall, he had to choose between two able people, for two positions, one a bit more prestigious than the other, and he wanted to be scrupulously fair about his assignments. So he considered flipping a coin, but then playfully worried about "a possible bias in the coin." At this point, Norman Johnson came to his "rescue," referring him to a result by von Neumann: flip the biased coin twice at a time until one gets a "HT" or "TH," two outcomes which can then be mapped into a mean $\frac{1}{2}$ Bernoulli random variable. Perhaps Wassily even made his appointments based on this technique; I suspect he did. But, of more interest, he raised the question of the efficiency of von Neumann's procedure for producing an unbiased coin: is it possible to produce a mean $\frac{1}{2}$ Bernoulli random variable from a sequence of independent Bernoulli random variables $X_1, \ldots, X_N$, where N is a stopping time of minimal expectation? The results of the subsequent research were interesting, and surprising, if not earth shattering. Later, Wassily described this research activity as one of his "most enjoyable."

The *Preface* to Wassily's *Collected Works* concludes with the following words:

> "A few months before Wassily's death, we visited him to request permission to produce this book. He expressed surprise (and pleasure) that the enterprise might be considered worthwhile. Then he offered us a drink. We asked what was available, he thought for a few moments, trying to recall the name of the liqueur (Benedictine), then said, 'Er... I forget. My memory is bad—but the liqueur is good.'
> 
> We shall treasure the opportunities we had to learn from Wassily and to appreciate his gentle humour."

## ACKNOWLEDGMENTS

This paper is based in large part on a memoir prepared for the U.S. National Academy of Sciences (Fisher and van Zwet, 2005). We are most grateful for their permission to reproduce most of that material, and to Professor Edward George, for encouragement and advice about modifying the article for publication in Statistical Science. We have been greatly assisted by many people in preparing the paper. In particular, Virginia Hoeffding went to a great deal of trouble to find a box of old papers and photographs from her uncle's [Wassily's] estate and make them available to us; Eugene Seneta provided a lot of help in interpreting these materials. We also acknowledge the considerable assistance of (the late) Stamatis Cambanis, Ildar Ibragimov, Ross Leadbetter, (the late) June Maxwell, Ingram Olkin and Gordon Simons.

NIF's research was supported by ValueMetrics Australia.